# Research on the Application of Large Language Models in Automatic Question Generation: A Case Study of ChatGLM in the Context of High School Information Technology Curriculum

Yanxin Chen[a,b], Ling He[a,b,*]

[a]Jiangxi Science and Technology Normal University, Nanchang, China; [b]VR Perception and Interaction Key Laboratory, Nanchang, China; *Corresponding Author

**Abstract**

This study investigates the application effectiveness of the Large Language Model (LLMs) ChatGLM in the automated generation of high school information technology exam questions. Through meticulously designed prompt engineering strategies, the model is guided to generate diverse questions, which are then comprehensively evaluated by domain experts. The evaluation dimensions include the Hitting (the degree of alignment with teaching content), Fitting (the degree of embodiment of core competencies), Clarity (the explicitness of question descriptions), and Willing to use (the teacher's willingness to use the question in teaching). The results indicate that ChatGLM outperforms human-generated questions in terms of clarity and teachers' willingness to use, although there is no significant difference in Hitting and Fitting. This finding suggests that ChatGLM has the potential to enhance the efficiency of question generation and alleviate the burden on teachers, providing a new perspective for the future development of educational assessment systems. Future research could explore further optimizations to the ChatGLM model to maintain high fit and hit rates while improving the clarity of questions and teachers' willingness to use them.

## 1 Introduce

### 1.1 BackGroud

High school information technology, as a required course for high school students, presents a rich and extensive coverage of knowledge points, imposing a significant teaching burden on information technology teachers. In recent years, the rise of Large Language Models (LLMs), particularly the application of ChatGLM, offers new possibilities for addressing this issue. Through carefully designed prompts, LLMs can generate questions that meet the requirements for high school information technology assessment, thereby providing a more efficient and convenient teaching tool for information technology teachers.

### 1.2 Problem Statement

This study is dedicated to exploring the application effects of LLMs, particularly ChatGLM, in the field of automatic question generation for high school information technology, aiming to assess whether they can rival human question setters. Through carefully designed prompt engineering strategies, the model is guided to generate diverse questions, which are then comprehensively evaluated by domain experts. The evaluation dimensions include key indicators such as the hit rate, relevance, clarity, and willingness of the questions, with the goal of verifying the practical application potential of LLMs in simulated question setting for high school information technology. This research not only contributes to enhancing the intelligence level of automatic question generation systems but also provides new perspectives and practical foundations for the future development of educational technology.

### 1.3 Objective

This study is committed to a thorough exploration of the performance of LLMs, particularly ChatGLM, in the context of automatic question generation tasks within the high school information technology domain, with the aim of assessing whether they can match the capabilities of human question setters. Utilizing prompt engineering strategies, the study guides LLMs to generate corresponding exam questions, which are then meticulously evaluated by domain experts across multiple dimensions including hit rate, relevance, clarity, and the experts' willingness to use these questions. Through a comparative analysis of the various indicators of the questions generated by the model, this study seeks to validate the application potential of LLMs in the realm of simulated question setting for high school information technology, with the hope of providing theoretical foundations and practical references for further advancing educational informatization.

## 2 Literature Review

### 2.1 The development of LLMs

The development trajectory of LLMs can be outlined through several key technological and application milestones. Initially, the concept of large-scale language models dates back to 2005, focusing primarily on machine translation using massive n-gram models[8]. Although these models were state-of-the-art at the time, they were subsequently supplanted by more sophisticated neural network models as technology advanced. In 2017, Vaswani et al. introduced the Transformer model, a novel network architecture entirely based on attention mechanisms, which abandoned the previously prevalent recurrent and convolutional structures[1]. The introduction of the Transformer model marked a significant turning point in deep learning for natural language processing, as it demonstrated superior performance on multiple machine translation tasks and significantly reduced training times. Following this, BERT and the GPT series models emerged in 2018 and 2019, respectively. BERT (Bidirectional Encoder Representations from Transformers) significantly enhanced performance across various NLP tasks by pre-training deep bidirectional representations from unlabeled text[3]. Meanwhile, Brown et al. proposed the GPT (Generative Pre-trained Transformer), which advanced the development of pre-trained language models through autoregressive language modeling[7]. By 2020, OpenAI released GPT-3, a language model with 175 billion parameters, which made significant strides in few-shot learning and achieved levels of performance competitive with human experts on certain tasks[9]. The success of GPT-3 not only showcased the formidable capabilities of large-scale models but also sparked extensive discussions on issues such as model scale, data quality, and bias. In recent years, with the increase in computational resources and optimization of algorithms, LLMs have been able to handle increasingly complex tasks such as cross-modal learning and multilingual processing [11]. Additionally, the applications of these models have expanded into various fields including healthcare, engineering, and social sciences[5]。In summary, the evolution of LLMs is a rapidly evolving process that involves algorithmic innovation, hardware advancement, and the expansion of application scenarios. Looking ahead, with further technological development and deeper application penetration, LLMs are poised to play a pivotal role in numerous domains.

### 2.2 Current Status of LLMs in Automatic Question Generation

2.3 The application of Large Language Models (LLMs) in the field of automated question generation can be explored from several key aspects. Firstly, Vishal Pallagani et al. proposed that the application of LLMs in the domain of automated planning can be extended to automated question generation, where a series of questions are generated to test learners' knowledge and skills[10]. Secondly, Jason Wei et al. introduced the chain-of-thought prompting method, which enhances the model's performance on complex reasoning tasks, particularly crucial in automated question generation, such as when solving mathematical problems, the model can demonstrate each step of the solution, aiding students in understanding complex concepts[4]. Furthermore, Long Ouyang et al. proposed fine-tuning based on human feedback, which can optimize the quality and relevance of questions by collecting feedback from teachers or experts, ensuring effective assessment of students' learning outcomes[6]. Lastly, Hugo Touvron et al. mentioned efficient foundational language models like LLaMA, which can train high-performance models without relying on proprietary datasets, suitable for automated question generation, producing high-quality, targeted questions[2]。In summary, the application of LLMs in automated question generation primarily relies on their powerful capabilities in planning, reasoning, and learning.

## 3 Method

### 3.1 Collect Questions

The content of the propositions is derived from the "Data and Information" (Data & Info) and "Algorithm Basics" (Algo) chapters in the compulsory textbook for the first year of high school. In the "Data and Information" chapter, the basic concepts, characteristics of data, and the process of transforming data into information are detailed, along with the significant role of information in modern society. This chapter spans 15 pages with a total of approximately 20,000 words. Following this, the "Algorithm Basics" chapter, which is 16 pages long with about 22,000 words, provides an in-depth yet accessible introduction to the fundamental principles, classifications, design methods of algorithms, and their applications in computer science and other fields.

The proposition methods are divided into two categories: manual proposition and ChatGLM proposition. Manual propositions are primarily based on post-class exercises and past years' selected questions, aiming to consolidate students' classroom knowledge and assess their understanding. On the other hand, ChatGLM propositions utilize artificial intelligence technology, generating questions through the natural language processing model ChatGLM. These propositions dynamically adjust the difficulty and type of questions based on students' learning progress and ability levels, providing a personalized learning experience. In the innovative practice of ChatGLM propositions, we adopt a Prompt engineering strategy that combines role-playing, tasks, and pseudocode. This strategy ensures the accuracy and pertinence of conveying the intention and requirements of the propositions to the ChatGLM model through the structured expression of pseudocode, enhancing the scientific nature of the propositions. By this strategy, we guide the ChatGLM model to deeply understand the educational objectives and students' needs, generating questions that meet educational standards and are highly personalized. In terms of the number of propositions, each chapter has 10 questions, totaling 20 questions for both chapters, thus equating to 20 manual propositions and 20 ChatGLM propositions.

### 3.2 Assessment

We invited five teachers with extensive teaching experience in the field of high school information technology to conduct a comprehensive and in-depth evaluation of a series of questions. The evaluation process covered several key dimensions, including the hitting of question knowledge points (the degree to which the tested knowledge points align with the content of the textbook), the fitting (the degree to which the questions reflect core competencies), clarity (whether the description of the questions is clear and unambiguous), and willing to use (whether teachers are willing to use the questions in their teaching). To ensure the objectivity and accuracy of the evaluation results, each dimension was scored on a scale of 1 to 5.

## 4 Results

We first examined the accuracy performance of manual propositions and ChatGLM propositions across different chapters and analyzed the differences between the two through multiple evaluation metrics. By analyzing Figure 1, we found that the average accuracy of manual propositions in the Data & Info chapter was 4.36 with a standard deviation of 0.56, and in the Algo chapter, it was 4.28 with a standard deviation of 0.61. In comparison, the average accuracy of ChatGLM propositions in the Data & Info chapter was 4.18 with a standard deviation of 0.60, and in the Algo chapter, it was 4.16 with a standard deviation of 0.55. These data indicate that the average accuracy of manual propositions was slightly higher than that of ChatGLM propositions in both chapters, although the difference was not substantial. The comparison of standard deviations shows that manual propositions were more stable in the Data & Info chapter, while ChatGLM propositions were more consistent in the Algo chapter.

Further analysis of the data in Table 2 and Figure 2 evaluated the performance of Human and ChatGLM on the three indicators of Fitting, Clarity, and Willingness. On the Fitting indicator, the average score for Human was 4.11 with a standard deviation of 0.567, while the average score for ChatGLM was slightly lower at 4.10 with a standard deviation of 0.461. This indicates that the performance of both was very close on this indicator. On the Clarity indicator, the average score for Human was 4.03 with a standard deviation of 0.502, while the ChatGLM model showed a higher average score of 4.14 with a standard deviation of 0.569, indicating that ChatGLM performed slightly better than Human in terms of Clarity. On the Willing indicator, the average score for manual proposers was 3.05 with a standard deviation of 0.539, while the average score for the ChatGLM model was significantly higher, reaching 3.98 with a standard deviation of 0.568. This result clearly demonstrates that ChatGLM significantly outperformed Human on the Willing indicator.

To explore the impact of different proposition methods on the variables of hit rate, fit, clarity, and willingness, we employed a one-way analysis of variance (ANOVA) using Welch's method. The results in Table 3 show that the p-values for Hitting ($p = 0.067$), Fitting ($p = 0.891$), and Clarity ($p = 0.149$) did not reach significance levels ($p < 0.05$), indicating that the mean differences in these variables between different proposition methods were not significant. However, the p-value for Willing was less than 0.001, significantly lower than the conventional threshold, and the average Willing score for ChatGLM (3.98) was significantly higher than that for Human (3.05), strongly suggesting that there were significant mean differences in Willing between different proposition methods. In summary, only the Willing variable showed significant mean differences between different proposition methods, while the hit rate, fit, and clarity variables did not show such differences. Through interviews, the teachers who participated in the scoring indicated that they were willing to adopt ChatGLM because it could alleviate the burden of teaching.

## 5 Discussion

This study compared the hit rate performance of manual propositions and ChatGLM propositions across different chapters and evaluated the three indicators of Fitting, Clarity, and Willing to reveal the similarities and differences in the quality of propositions between the two methods. Although manual propositions had a slightly higher average accuracy than ChatGLM propositions in the Data & Info and Algo chapters, the difference was not significant. In the comparison of standard deviations, manual propositions were more stable in the Data & Info chapter, while ChatGLM propositions were more consistent in the Algo chapter. On the Fitting indicator, the performance of manual propositions and ChatGLM propositions was very close, showing similar levels of alignment between the proposition content and teaching objectives. However, on the Clarity indicator, ChatGLM propositions demonstrated a slightly better performance than manual propositions, which could be attributed to the linguistic generation advantages of ChatGLM, providing clearer and more accurate expressions. The most significant difference was observed in the Willing indicator, where the average score of ChatGLM propositions was significantly higher than that of manual propositions, indicating that teachers and evaluators were more inclined to accept and use propositions generated by ChatGLM. This result may reflect the potential value of ChatGLM in reducing the workload of teachers and improving the efficiency of proposition generation. Through one-way analysis of variance (ANOVA) using Welch's method, we found that only the Willingness variable showed significant mean differences between different proposition methods, while the Hitting, Fitting and Clarity variables did not exhibit such differences. This result further supports the advantage of ChatGLM in enhancing the acceptance of propositions.

**6 Conclusion**

This study conducted a detailed data analysis and statistical testing to compare the performance of manual propositions and ChatGLM propositions across multiple evaluation metrics, including accuracy, fit, clarity, and willingness. The results indicated that although manual propositions had a slightly higher average accuracy than ChatGLM propositions in certain chapters, the differences were not statistically significant. In the comparison of standard deviations, manual propositions exhibited greater stability in the Data & Info chapter, while ChatGLM propositions demonstrated more consistency in the Algo chapter. On the Fitting indicator, the performance of manual propositions and ChatGLM propositions was remarkably close, indicating similar levels of alignment between the proposition content and teaching objectives. However, on the Clarity indicator, ChatGLM propositions outperformed manual propositions slightly, likely due to the linguistic generation advantages of ChatGLM, which provided clearer and more accurate expressions. The most significant difference was observed in the Willing indicator, where the average score of ChatGLM propositions was significantly higher than that of manual propositions, suggesting that teachers and evaluators were more inclined to accept and use propositions generated by ChatGLM. This result may reflect the potential value of ChatGLM in reducing the workload of teachers and enhancing the efficiency of proposition generation. Through one-way analysis of variance (ANOVA) using Welch's method, we found that only the Willing variable showed significant mean differences between different proposition methods, while the Hitting, Fitting and Clarity variables did not exhibit such differences. This result further supports the advantage of ChatGLM in enhancing the acceptance of propositions.

In summary, ChatGLM demonstrates significant potential in improving the clarity of propositions and enhancing teachers' willingness to accept them. Although manual propositions still hold certain advantages in some aspects, the integration and optimization of ChatGLM could bring revolutionary changes to the educational assessment system. Future research could explore further optimizations of the ChatGLM model to enhance the clarity and teachers' willingness to use propositions while maintaining high levels of fit and accuracy. Considering the continuous technological advancements, future educational assessment systems are likely to integrate more AI tools similar to ChatGLM to improve teaching efficiency and assessment quality.

## 8 Appendix

**Table 1 Prompt Details**

| |
|---|
| Role: You are a high school information technology teacher, possessing highly specialized subject knowledge and pedagogical knowledge, and you can understand Chinese accurately and without error.<br>Task: Create a written exam question for students.<br>Requirements: Closely align with the content of the textbook.<br>Number of questions: 1.<br>Textbook content: {textbook_content}<br>Output content: Output the content in the format of pseudocode logic.<br>[<br>    {"Seq":XX, Question": "XXX"}<br>] |

**Table 2 Statistics from Different Methods**

| | Method | Mean | SD |
|---|---|---|---|
| Hitting | Human | 4.32 | 0.584 |
| | GLM | 4.17 | 0.570 |
| Fitting | Human | 4.11 | 0.567 |
| | GLM | 4.10 | 0.461 |
| Clarity | Human | 4.03 | 0.502 |
| | GLM | 4.14 | 0.569 |
| Willing | Human | 3.05 | 0.539 |
| | GLM | 3.98 | 0.568 |

*Note: GLM=ChatGLM

**Table 3. One-way ANOVA (Welch's) Results for the Effect of Sources**

| | df1 | df2 | p |
|---|---|---|---|
| Hitting | 1 | 198 | 0.067 |
| Fitting | 1 | 190 | 0.891 |
| Clarity | 1 | 195 | 0.149 |
| Willing | 1 | 197 | < .001 |

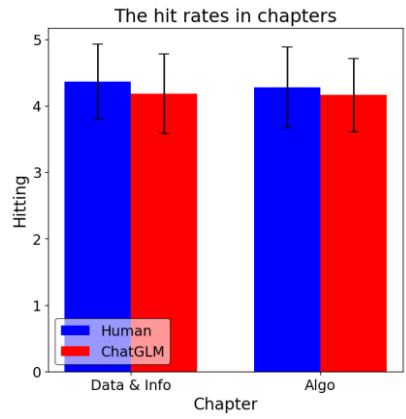

Figure 1 Hitting in Chapters

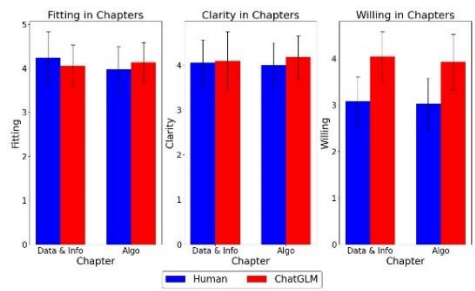

Figure 2 Fitting, Clarity and Willing